\documentclass[letter, traditabstract]{aa} 
\usepackage{graphicx}
\usepackage{txfonts}
\usepackage{natbib}

\newcommand{\kms}{km.s$^{-1}$}
\newcommand{\kmss}{km.s$^{-1}\;$}

\newcommand{\vsinis}{$v\sin i\;$}

\newcommand{\ang}{\AA}

\def\gtrsim{\mathrel{\hbox{\rlap{\hbox{\lower4pt\hbox{$\sim$}}}\hbox{$>$}}}}
\def\ltsim{\mathrel{\hbox{\rlap{\hbox{\lower4pt\hbox{$\sim$}}}\hbox{$<$}}}}
\newcommand{\degre}{$^{\circ}$}
\newcommand{\degres}{$^{\circ}\;$}
\newcommand{\bft}{}

\begin{document}

\titlerunning{First HARPSpol discoveries of magnetic fields in massive stars}
\title{First HARPSpol discoveries of magnetic fields in massive stars
\thanks{Based on observations collected at the European Southern Observatory, Chile (Program ID 187.D-0917)}
}

\author{E. Alecian
          \inst{1}
          \and
          O. Kochukhov \inst{2}
          \and
          C. Neiner \inst{1}
          \and
          G.A. Wade \inst{3}
          \and
          B. de Batz \inst{1}
          \and
          H. Henrichs \inst{4}
          \and
          J.H. Grunhut \inst{3}
          \and
          \\J.-C. Bouret \inst{5}
          \and
          M. Briquet \inst{6,1,8}
          \and
          M. Gagne \inst{7}
          \and
          Y. Naze \inst{8}
          \and
          M.E. Oksala \inst{9}
          \and
          T. Rivinius \inst{10}
          \and
          \\R.H.D. Townsend \inst{11}
          \and
          N.R. Walborn \inst{12}
          \and
          W. Weiss \inst{13}
          \and
          the MiMeS collaboration
          }

\institute{LESIA-Observatoire de Paris, CNRS, UPMC Univ., Univ. Paris-Diderot, 5 place Jules Janssen, F-92195 Meudon Principal Cedex, France,
              \email{evelyne.alecian@obspm.fr}
              \and
              Department of Physics and Astronomy, Uppsala University, Box 516, SE-751 20 Uppsala, Sweden
              \and
              Dept. of Physics, Royal Military College of Canada, PO Box 17000, Stn Forces, Kingston K7K 7B4, Canada
              \and
              Astronomical Institute Anton Pannekoek, University of Amsterdam, Science Park 904, 1098XH Amsterdam, Netherlands
              \and
              Laboratoire d'Astrophysique de Marseille, Traverse du Siphon, BP8-13376 Marseille Cedex 12, France 
              \and
              Instituut voor Sterrenkunde, Katholieke Universiteit Leuven, Celestijnenlaan 200 D, 3001 Leuven, Belgium
              \and
              Department of Geology and Astronomy, West Chester University, West Chester, PA 19383
              \and
              FNRS-Institut d'Astrophysique et de G\'eophysique, Universit\'e de Li\`ege, All\'ee du 6 Ao\^ut 17, Bat B5c, B 4000 - Li\`ege, Belgium
              \and
              Department of Physics and Astronomy, University of Delaware, Newark, DE 19716, USA
              \and
              European Organisation for Astronomical Research in the Southern Hemisphere, Casilla 19001, Santiago 19, Chile
              \and
              Dept. of Astronomy, University of Wisconsin-Madison, 475 N. Charter Street, Madison WI 53706-1582, USA
              \and
              Space Telescope Science Institute, 3700 San Martin Drive, Baltimore, MD 21218, USA
              \and
              Institut f\"ur Astronomie, Universit\"at Wien, T\"urkenschanzstrasse 17, A-1180 Wien, Austria
             }

   \date{Received September 15, 1996; accepted March 16, 1997}

 
  \abstract
{\bft In the framework of the Magnetism in Massive Stars (MiMeS) project, a HARPSpol Large Program at the 3.6m-ESO telescope has recently started to collect high-resolution spectropolarimetric data of a large number of Southern massive OB stars in the field of the Galaxy and in many young clusters and associations. 
In this Letter, we report on the first discoveries of magnetic fields in two massive stars with HARPSpol - HD 130807 and HD 122451, and confirm the presence of a magnetic field at the surface of HD 105382 that was previously observed with a low spectral resolution device. The longitudinal magnetic field measurements are strongly varying for HD 130807 from $\sim$-100~G to $\sim$700 ~G. Those of HD 122451 and  HD 105382 are less variable with values ranging from $\sim$-40 to -80~G, and from $\sim$-300 to -600~G, respectively. The discovery and confirmation of three new magnetic massive stars, including at least two He-weak stars, is an important contribution to one of the MiMeS objectives: the understanding of origin of magnetic fields in massive stars and their impacts on stellar structure and evolution.}
   \keywords{Stars: massive -- Stars: magnetic field -- Stars: chemically peculiar -- Stars: individual: HD~122451, HD~105382, HD~130807}

   \maketitle
%

\section{Introduction}

MiMeS\footnote{http://www.physics.queensu.ca/$\sim$wade/mimes} (Magnetism in Massive Stars) is a large collaboration that aims to address many issues concerning the magnetism of massive stars. One goal in particular is to determine the global magnetic properties of massive stars with the help of Large Programs (LP)  that have been allocated on the high-efficiency high-resolution spectropolarimeters ESPaDOnS (Canada France Hawaii Telescope, Hawaii) and Narval (Telescope Bernard Lyot, France). These programs aim to observe about 200 massive OB field stars (the Survey Component or SC), in order to search for magnetic fields, {\bft confirm those previously suspected,} and derive statistical properties {\bft \citep{wade09,grunhut11b}}. They also aim to observe intensely about 30 already known magnetic massive stars (the Targeted Component {\bft or} TC) in order to map in detail their surface magnetic fields. This smaller sample of stars is dedicated to the study of the interplay of magnetic fields with the stellar structure, environment and evolution at high mass \citep[e.g.][]{grunhut09,oksala11}.

In 2010, the polarimeter HARPSpol was commissioned at the 3.6m-ESO telescope (La Silla, Chile). For the first time, we can access the Southern hemisphere with data quality similar to ESPaDOnS and Narval. Therefore, a Large Program was established to complete the ESPaDOnS/Narval field sample, and to take the first steps toward observing massive stars in various open clusters {\bft and associations} of different ages, to investigate the magnetic field evolution, and the impact of magnetic fields on stellar evolution.

The HARPSpol sample is divided in two components (SC and TC)  to follow the same strategy as the Narval and ESPaDOnS LPs. The HARPSpol SC sample contains about 180 stars including $\sim$110 stars in 7 clusters, and $\sim$70 stars in the field of the Galaxy. The former have been selected from the Catalogue of Open Cluster Data \citep{kharchenko05}, while the latter have been chosen from the International Ultraviolet Explorer (IUE) data archive but also from other catalogues or publications containing highly probable magnetic stars, in accordance with the ESPaDOnS/Narval target selection \citep[][]{wade09}.

This HARPSpol Large Program {\bft was} allocated four separate runs over two years. During the first run in May 2011, 57 stars were observed including one magnetic calibrator and one TC target for which the results will be presented in a forthcoming paper. In this letter, we report on the first discoveries of magnetic fields in massive stars with HARPSpol in HD~130807 and HD~122451, and we confirm the magnetic field in HD~105382 previously detected with the low-resolution spectropolarimeter FORS 1 \citep{kochukhov06,hubrig06}. Among the 55 SC stars observed during this run, those three stars are the only ones in which a magnetic field was detected. In Section 2, we present the observations and reduction techniques. In Section 3, we detail the HARPSpol results on each star, and discuss them in Section 4.

%

\section{Observations}

\begin{table}[t]
\caption{Log of observations. Columns 1 and 2 give the date, Universal Time (UT) and Heliocentric Julian Date of the observations. Columns 3 and 4 give the total exposure time and the number of polarimetric sequences. Columns 5 and 6 give the peak S/N per CCD pixel (at $\sim$501~nm for HD~130807 and at $\sim$518~nm for HD~122451 and HD~105382) in the spectra, and the S/N per 1.4~\kmss (for HD 130807 and HD 105382) and 4.2~\kmss (for HD 122451) pixels in the LSD Stokes $V$ profiles. Columns 7, 8 and 9 give the longitudinal magnetic field, the magnetic detection probability, and the detection type (see text).}
\label{tab:log}
\centering
\begin{tabular}{@{}l@{}c@{}c@{\,\,}c@{\,\,}r@{\,\,}r@{\,\,}r@{$\pm$}l@{\,\,}c@{\,\,}l@{}}
\hline\hline
Date (d/m)       & HJD               & $t_{\rm exp}$ & \# & S/N & S/N     & \multicolumn{2}{c}{$B_{\ell}$} & $P_{\rm det}$ & \\
UT                   & (2 455 000+) & (s)                  &        &      & (LSD) & \multicolumn{2}{c}{(G)}           &                        & \\
\hline
\multicolumn{9}{c}{\bft HD 130807} \\
23/05 05:23 & 704.724 & 4000 & 1 & 500 & 4600 & 292 & 26  & 1.00000 & DD \\
27/05 06:06 & 708.754 & 1200 & 1 & 520 & 4800 & -94 & 26 & 1.00000 & DD \\
28/05 05:58 & 709.748 & 6000 & 2 & 630 & 5700 & 677 & 21   & 1.00000 & DD \\
\\[-5pt]
\multicolumn{9}{c}{\bft HD 122451} \\
23/05 04:16 & 704.677 & 1340 &   5 & 1680 & 33200 & -43 & 20\tablefootmark{\bft a} & 1.00000 & DD \\
27/05 05:21 & 708.723 & 1200 & 20 & 3000 & 61800 & -83 & 14\tablefootmark{\bft a} & 1.00000 & DD \\
28/05 04:34 & 709.690 & 1680 & 14 & 1260 & 26000 & -66 & 29\tablefootmark{\bft a} & 0.98173 & ND\\
\\[-5pt]
\multicolumn{9}{c}{\bft HD 105382} \\
25/05 01:19 & 706.555 & 3200 & 1 & 990 & 9650 & -622 & 26   & 1.00000 & DD \\
26/05 01:59 & 707.582 & 4000 & 1 & 900 & 8840 & -298 & 32   & 1.00000 & DD \\
28/05 00:10 & 709.506 & 4800 & 1 & 920 & 8900 & -406 & 32   & 1.00000 & DD \\
\hline
\end{tabular}
\tablefoot{
\tablefoottext{\bft a}{\bft Those values are for  HD 122451 B only.}
}
\end{table}

We used the HARPSpol polarimeter \citep{piskunov11}, combined with the HARPS spectrograph \citep{mayor03}, installed at the 3.6m ESO telescope at La Silla Observatory (Chile), yielding spectra with resolving power $\lambda / \Delta \lambda$ of about $105\,000$, and covering the 380--690~nm wavelength region. All spectra were recorded as sequences of 4 individual sub-exposures taken in different configurations of the polarimeter, in order to yield a full circular polarisation analysis, as described by \citet{donati97}. The data were reduced using the package ``REDUCE'' described by \citet{piskunov02}. After reduction, we obtained the intensity Stokes $I$ and the circular polarisation Stokes $V$ spectra of the stars, both normalised to the continuum. A null spectrum ($N$) was also computed in order to diagnose spurious polarisation signatures, and to help to verify that the signatures in the Stokes $V$ spectrum are of stellar origin. The log of the observations is presented in Table 1.

To increase the effective signal to noise ratio (S/N) of our data, we applied the Least Squares Deconvolution \citep[LSD;][]{donati97} procedure using tailored line masks of appropriate temperature and gravity for each star. The masks were first computed using Kurucz ATLAS 9 models of solar abundance \citep{kurucz93}, with intrinsic line depths larger than 0.1. We then excluded from these masks {\bft hydrogen} Balmer lines, and lines whose Land\'e factor is unknown. Finally we have modified the line depths {\bft to} take into account the relative depth of the lines of the observed
spectra. The resulting masks contain 394, 592, and 394 lines for HD 1030807, HD 122451 and HD 105382, respectively. The S/N of the LSD Stokes $V$ profiles is about 10 times larger than the S/N in the original spectra (Table \ref{tab:log}).

In order to perform a reliable magnetic field diagnosis, we have computed the detection probability inside the LSD $V$ profiles \citep[as described in ][]{donati97}. We consider that an observation displays a ``definite detection" (DD) {\bft of Stokes $V$ Zeeman signature} if the probability is larger than 0.99999, a ``marginal detection" (MD) if it falls between 0.999 and 0.99999, and a ``null detection" (ND) otherwise (see Table 1). All observations of HD 130807 and HD 105382 display DD while two DD and one ND have been obtained for HD 122451. The LSD $I$, $V$, and $N$ profiles are plotted in Fig. \ref{fig:lsd}. In almost all of our observations Zeeman signatures, as broad as the $I$ profiles, are clearly detected in the $V$ profiles, while the $N$ profiles are consistent with the noise. These results allow us to confidently affirm that magnetic fields are present at the surface of these stars. 

We measured for each observation the line-of-sight component of the magnetic field averaged over the visible stellar surface (the so-called longitudinal magnetic field or $B_{\ell}$), by integrating the $I$ and $V$ profiles over the ranges $[-50,60]$, $[-80,100]$, and $[-70,105]$ \kms for HD 130807, HD 122451, and HD 105382, respectively  \citep[as described by ][]{alecian09}. The values are reported in Table 1.

%

\section{Results}

\subsection{HD 130807}

\begin{figure*}
\centering
\includegraphics[width=6cm,clip=true]{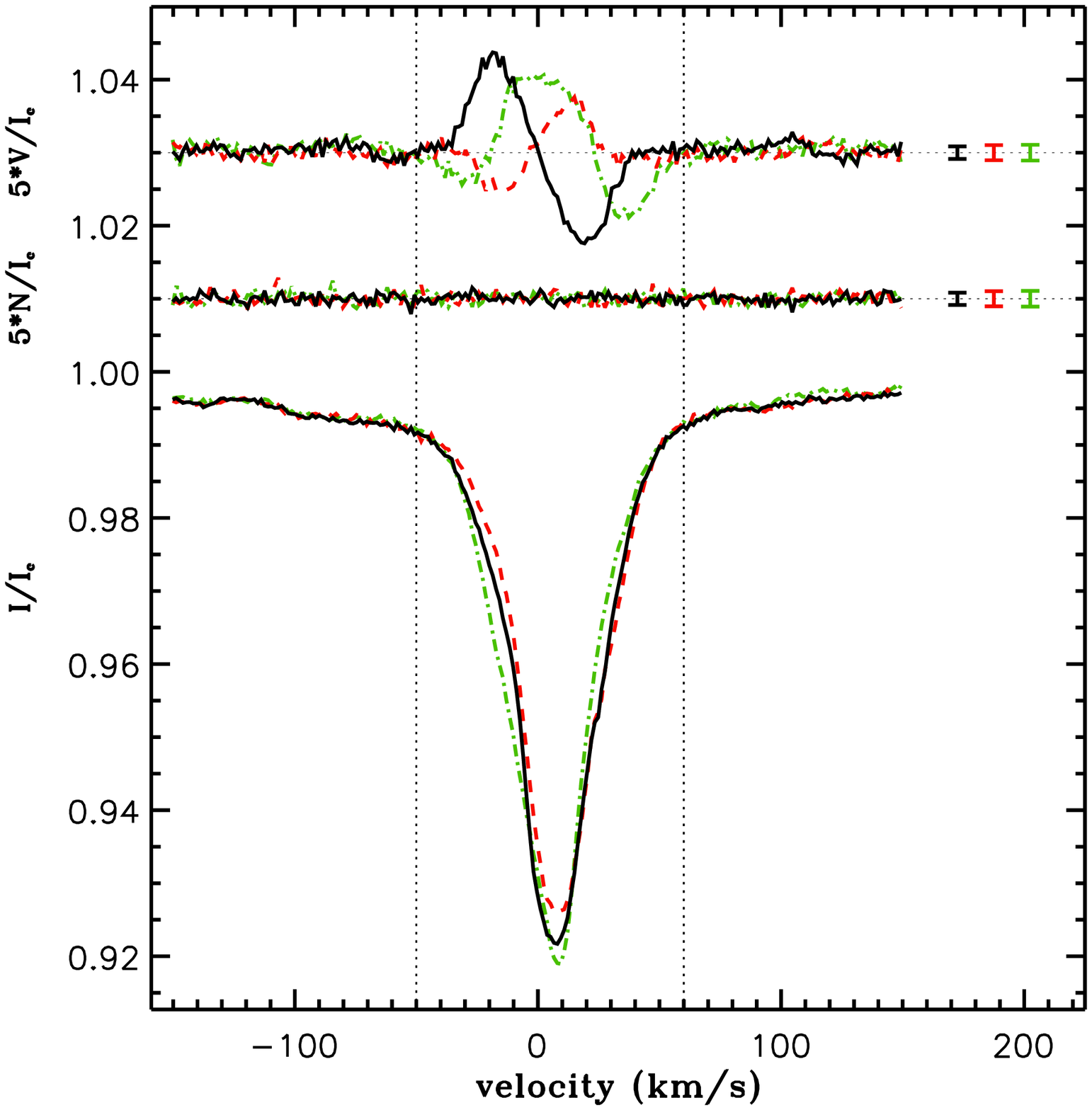}
\includegraphics[width=6cm,clip=true]{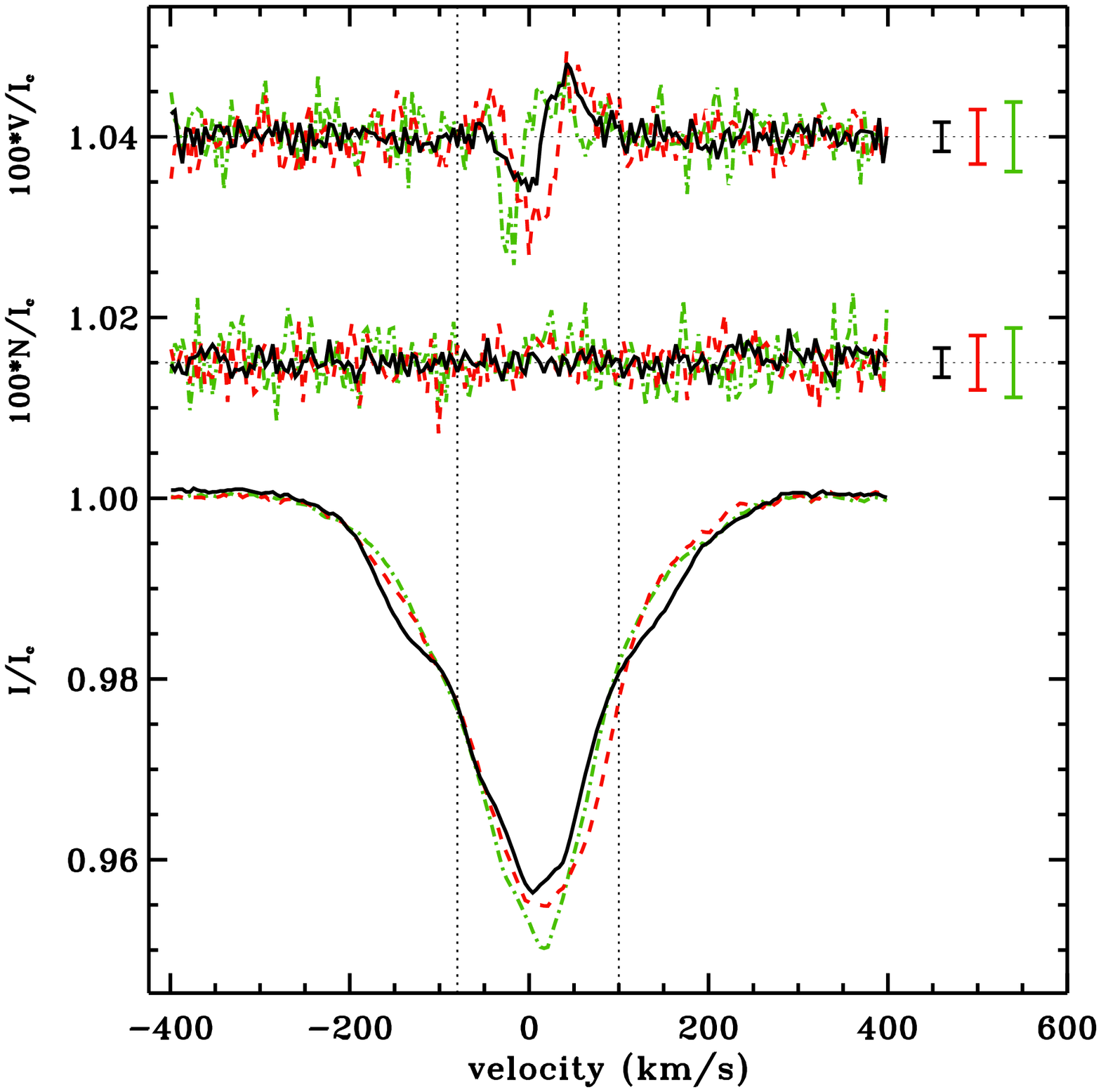}
\includegraphics[width=6cm,clip=true]{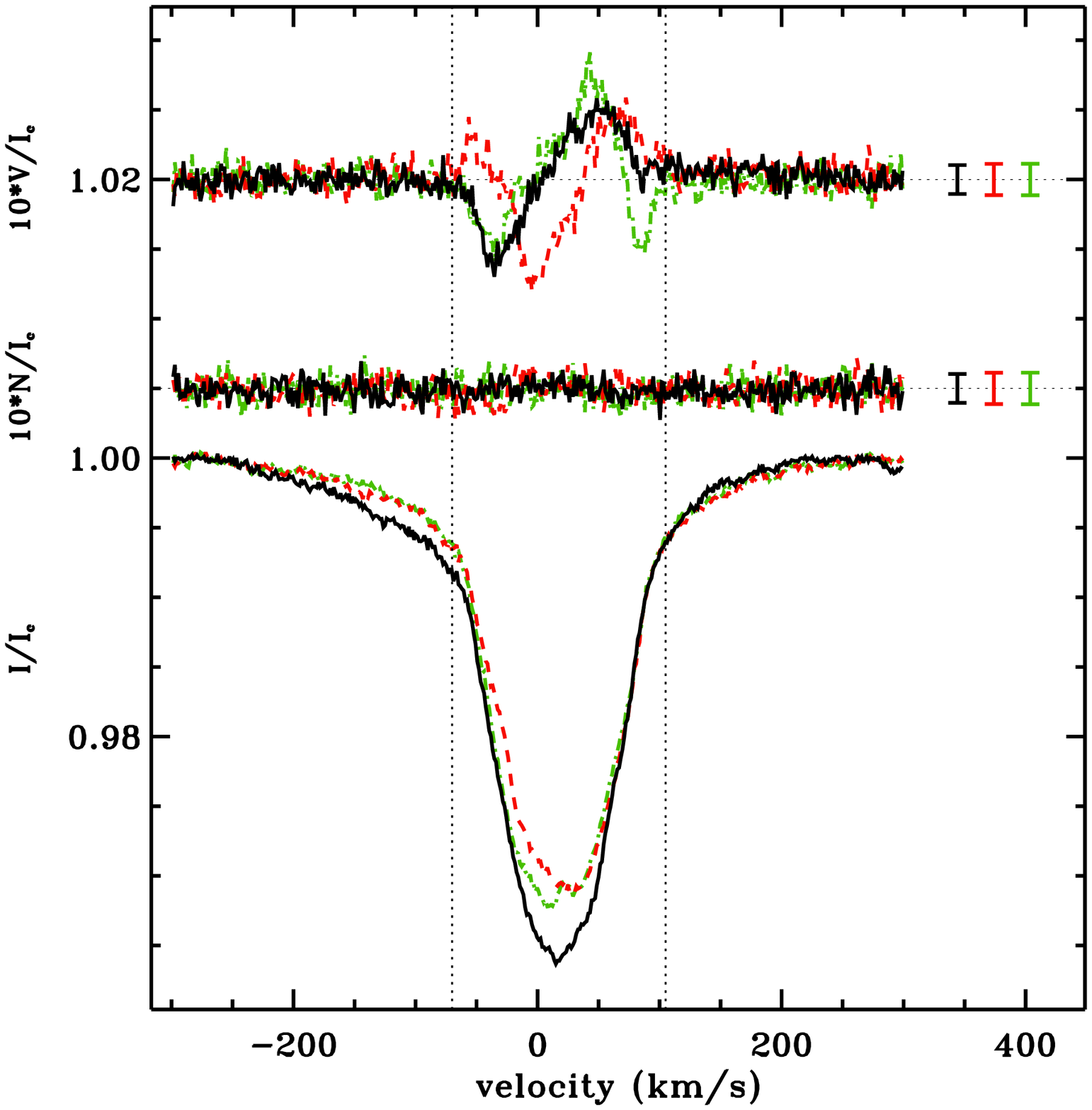}
\caption{LSD $V$ (top), $N$ (middle), and $I$ (bottom) profiles of HD~130807 ({\it left}), HD~122451 ({\it middle}), and HD~105382 ({\it right}). The mean error bars are plotted next to each profile. The $V$ and $N$ profiles have been shifted and amplified for display purpose. The dotted vertical lines indicate the integration ranges for the calculation of $B_{\ell}$. {\it left}: full black line: 28/05, red-dashed line: 27/05, green dot-dashed line: 23/05. {\it middle}: full black line: 27/05, red-dashed line: 23/05, green dot-dashed line: 28/05. {\it right}: full black line: 25/05, red-dashed line: 28/05, green dot-dashed line: 26/05.}
\label{fig:lsd}%
\end{figure*}

{\bft  HD 130807 ($o$ Lup) is member of the Sco-Cen association \citep{kharchenko05}. A companion was detected at an angular distance varying from 0.07 to 0.14 arcsec \citep{perryman97,mcalister90}. 
According to the angular separation, the light of both components entered the HARPSpol fibers during our observations of this target.}

\begin{figure}
\centering
\includegraphics[width=9cm,clip=true]{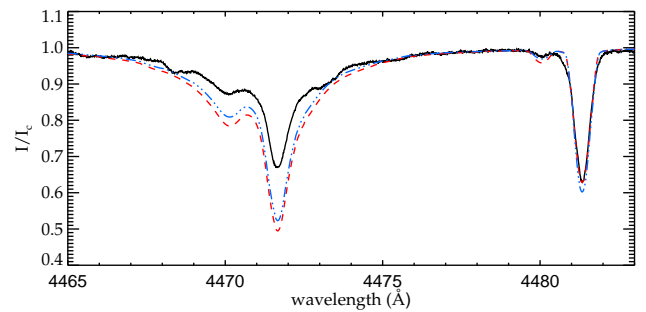}
\caption{Spectrum (full black) of HD 130807 plotted around He~{\sc i}~4471~\ang\ and Mg~{\sc ii} 4481 \ang. Synthetic spectra of 18000~K (dashed red) and 17000~K (dot-dot-dot-dashed blue) are overplotted.}
\label{fig:sphd130807}%
\end{figure}

{\bft From a visual inspection} we find that most of the spectrum of HD 130807 is consistent with a synthetic spectrum of a single star of {\bft effective temperature} $T_{\rm eff}=18000$~K, {\bft surface gravity} $\log g=4.25$ (cgs), broadened by $v\sin i=25$~\kms, calculated using TLUSTY non-LTE atmosphere models and the SYNSPEC code \citep{hubeny88,hubeny92}. However, we observe that all He~{\sc i} lines are substantially weaker than the synthetic ones calculated with solar abundance (Fig. \ref{fig:sphd130807}), while the Si~{\sc ii} lines are considerably stronger. The Si, N and Fe lines show variability in depth and shape on a timescale of 1~d. These characteristics suggest that HD~130807 is an He-weak star with abundance spots on its surface \citep{jaschek74}.

Magnetic signatures are detected in almost all the lines of the spectrum, similar to the LSD one (Fig. \ref{fig:lsd} {\it left}). Many additional lines are observed in the spectrum that could be due to Fe~{\sc ii}, Fe~{\sc iii}, or Ti~{\sc ii} enhancements. All these lines show Zeeman signatures similar to the others, with the same variations from one night to the other. They can therefore be attributed to the same star, rather than a companion, and they are probably the result of the chemical peculiarities at the surface of the star.

A significant shift in radial velocity ($\sim 6$ km/s) is detected in the strongest spectral lines including Balmer lines, between May 22 and May 26-27. The maximum reported angular separation between both visual components implies a distance $\ge17$~AU, and therefore a period $\ge27$ years. This radial velocity shift cannot therefore be due to the reported visual companion. A third companion very close to the primary could explain these variations, but more observations are required to fully understand all the peculiarities observed in the spectrum.

The variations observed in the $V$ profiles over 6 days (Fig. \ref{fig:lsd} {\it left}) can be understood in terms of the oblique rotator (OR) model that consists of an inclined dipole placed inside a rotating star \citep{stibbs50}. The rotational modulation of the shape of the $V$ profiles and of the $B_{\ell}$ values that vary from {\bft -94 to 677~G} (Table 1) suggest that the rotation period of the star should be between 1 and 6 days. More observations, well sampled over the rotation period, are required in order to fully characterise the magnetic field and better constrain the period of HD 130807.

\subsection{HD 122451}

{\bft HD 122451 ($\beta$ Cen) is a double-lined spectroscopic binary with components of similar effective temperatures (25000 K) and gravities ($\log g=3.5$, cgs), and a $\beta$~Cep-type pulsating primary. The system is highly eccentric ($e = 0.835$) and orbits with a period of 357 days \citep{ausseloos02,ausseloos06,davis05}.}

We obtained {\bft three} observations of HD~122451 during 3 different nights. In order to avoid potential false magnetic detections due to pulsations, we have split each observation into many sequences of four spectra (Table \ref{tab:log}), so that the exposure time for one sequence is much shorter than the pulsation period.

The spectra of HD 122451 clearly show two components of similar temperatures but different broadening, confirming the SB2 nature of the system. We adopt the same definition as \citet{ausseloos06} for the primary and secondary, i.e. as the broad-line and narrow-line components, respectively. In order to measure the \vsinis of both stars, we performed a least-squares fit to few individual spectral lines with the sum of two functions calculated as the convolution of a Gaussian of instrumental width and a rotation function as described by \citet{gray92} \citep[see details of the fitting procedure {\bft used by}][]{alecian08}. We find a \vsinis of $190\pm20$~\kmss and $75\pm15$~\kmss for the primary and secondary, respectively. When compared with TLUSTY/SYNSPEC synthetic spectra our observations are consistent with $T_{\rm eff}=25000$~K and $\log g=3.5$~(cgs), in agreement with the work of \citet{ausseloos06}. The spectral lines appear distorted and show rapid variations very likely due to $\beta$~Cep-type pulsations. No obvious abundance peculiarity, nor manifestation of circumstellar matter is observed within the spectra.

In Fig. \ref{fig:lsd} ({\it middle}), we superimposed the LSD $I$, $V$, and $N$ profiles of our observations. According to the ephemeris of \citet{ausseloos06}, the 3 observations are roughly at the same orbital phase ($\sim$0.5), and both components have {\bft similar} radial velocities ($\sim$14 and $\sim$4~\kmss for the primary and secondary respectively), which explains why it is difficult to distinguish both components in the profiles. The shape of the LSD $I$ profile shows variations during the run that can be understood in terms of radial pulsations in the primary, which would occasionally broaden the profile. As a result, both components can be clearly distinguished in the profile of May 27 (full black line in Fig. \ref{fig:lsd} {\it middle}), while it is less obvious in the other observations.

Zeeman signatures are detected in many individual spectral lines, {\bft as in} the LSD $V$ profiles. The signatures are as broad as the secondary profile, meaning that the magnetic field is detected only in the secondary component of the system. However, considering the faint Zeeman signatures in the secondary, and the broad line shape of the primary, a magnetic field of the same strength as the secondary's could exist in the primary, without being detected in our observations. {\bft In order to estimate the $B_{\ell}$ values of the secondary, we need to extract the $V$ and $I$ profiles of the secondary only, from the profiles of the binary. Without any evidence of a magnetic field in the primary, we neglect its contribution to the $V$ profile, which we consider to be entirely from the secondary. On the contrary, the $I$ profile needs to be corrected. With} this aim, we have first fitted the $I$ profile of the binary with the method described above for the individual spectral lines. Then we have subtracted from the observed $I$ profile the fit of the primary. Finally we have measured $B_{\ell}$ using the corrected $I$ and the original $V$ profiles. The values are reported in Table 1.

\subsection{HD 105382}

HD 105382 (= HR 4618) is member of the Sco-Cen association \citep{kharchenko05}. \citet{briquet04} classified it as He-weak with He patches enhanced where Si is depleted, and derived a $T_{\rm eff}$ of $17400\pm800$~K, a $\log g=4.18\pm0.20$ (cgs), a rotation period of $1.295 \pm 0.001$~d and an inclination angle {\bft of the rotation axis to the line of sight} $i=50\pm10$\degre.

In our {\bft three} spectra, we observe strong variations in the spectral lines, mainly in He~{\sc i}, Si~{\sc ii} and Fe~{\sc iii}, that are due to abundance spots on the stellar surface described by \citet{briquet04}. Clear Zeeman signatures are detected in the metallic and Balmer lines, as well as in the LSD $V$ profiles (Fig. \ref{fig:lsd} {\it right}). The rotation phases of our observations, calculated with a rotation period of 1.295~d, are very different (0.35, 0.14, and 0.63), and yet the $V$ profiles are all similarly negative (Table 1). According to the OR model, this implies that the magnetic obliquity angle (with respect to the rotation axis) cannot be very high ($|\beta| < 40^{\circ}$ if $i = 50^{\circ}$), otherwise the positive magnetic pole would sometimes appear on the visible stellar hemisphere, creating a positive profile at least once during the run.

HD 105382 was independently discovered as magnetic by \citet{kochukhov06} and \citet{hubrig06}. \citet{briquet07} derived the longitudinal field from FORS~1 observations and found values ranging from $-923$~G to $840$~G. Among their four values, the May 2004 {\bft observation} ($840\pm58$~G) is clearly inconsistent with our data as positive values are not expected. \citet{bagnulo11} very carefully re-reduced the same FORS~1 data and {\bft found values consistent with those of Briquet et al. (2007)} except for that observation ({\bft for which they derived} $B_{\ell}=-29\pm69$~G).

We performed a least-square sinusoidal fit to our $B_{\ell}$ values simultaneously with the Briquet et al. (2007) data and could find a solution only by removing the May 2004 datapoint. We also performed an independent fit using the re-reduced data of \citet{bagnulo11} and found a similar result. In both cases, the derived period is consistent with that of Briquet et al. (2004). The fitted $B_{\ell}$ values are very similar in both cases, {\bft varying} from $-670$~G to $-20$~G, {\bft and} implying a magnetic obliquity of $\sim38$\degres and a polar field strength of $\sim2.3$~kG, assuming a dipole field \citep{borra80}.

%
\section{Discussion}

We report direct detections of magnetic fields in three hot B-type stars (18000 K - 25000 K), among a sample of 55 stars in which we were searching for magnetic fields with HARPSpol. Two of them (HD 122451 and HD 130807) are completely new detections. For the other one - HD 105382 {\bft - this} is the first direct detection of a Zeeman signature. One of the main MiMeS results is the systematic detection of chemical peculiarities at the surface of magnetic hot stars \citep[and conversely, e.g.][]{grunhut11a}. Among the three stars discussed in this paper two are unambiguously He-weak. The third one (HD 122451) belongs to a binary system with a $\beta$~Cep primary, that makes the interpretation of the spectrum and the detection of peculiarities inside spectral lines very difficult. More observations well sampled over the orbital period of the system are required in order to first confirm a magnetic detection in only the secondary, and then disentangle the pulsation and chemical peculiarity effects.

The interplay between radiative and magnetic forces \citep{hunger99} is usually assumed to be at the origin of the over- or under-abundant He spots at the surface of hot magnetic B stars. These spots are very often correlated with the stellar magnetic fields \citep[e.g.][]{veto90}. In the case of HD~105382, \citet{briquet04} found a large He spot at a latitude of 60\degre. If our estimate of its magnetic obliquity is confirmed, this spot would be situated close to the South magnetic pole, demonstrating once more the importance of magnetic fields in the formation of chemical spots.

More observations of these magnetic stars are planned within the HARPSpol large program in order to perform detailed mapping of their magnetic fields, and better confront the models and theories of magnetic massive stars.

\begin{acknowledgements}
We wish to thank the Programme National de Physique Stellaire (PNPS) for their support. JHG and GAW acknowledges support from NSERC. M.B. acknowledges the Fund for Scientific Research -- Flanders for a grant for a long stay abroad, and she is a F.R.S.-FNRS Postdoctoral Researcher. RHDT acknowledges support from NSF grant AST-0908688. This research has made use of the SIMBAD database and the VizieR catalogue access tool, operated at CDS, Strasbourg (France), of INES data from the IUE satellite and of NASAs Astrophysics Data System.
\end{acknowledgements}

\bibliographystyle{aa}
\bibliography{harps_letter}

\end{document}